\documentclass[11pt,english,american,onecolumn]{article} 
\usepackage{bm}%
\usepackage[normalem]{ulem}
\usepackage[colorlinks=true,linkcolor=blue,citecolor=black]{hyperref}%
\usepackage{color}%
\usepackage[T1]{fontenc}
\usepackage[utf8]{inputenc}
\usepackage{algorithm}
\usepackage[font={small}]{caption}
\usepackage{authblk}
\usepackage{float}
\usepackage{mathtools}
\usepackage{amsmath}
\usepackage{amsthm}
\usepackage{amssymb}
\usepackage{bm}
\usepackage{physics}
\usepackage{graphicx}
\usepackage{color}
\usepackage{graphicx,epstopdf}
\usepackage{algpseudocode}
\usepackage{enumerate}
\usepackage{etoolbox}
\usepackage{lmodern}
\usepackage{stfloats}
\usepackage[usenames,dvipsnames]{xcolor}
\usepackage[utf8]{inputenc}
\usepackage{amsfonts}
\usepackage{comment}
\usepackage{hyperref}
\usepackage{authblk}
\usepackage[normalem]{ulem}
\usepackage{bm}
\usepackage{subfig}
\usepackage[left=1.8cm,right=1.8cm,top=1in,bottom=1in]{geometry}
\makeatother

\floatstyle{ruled}
\newfloat{algorithm}{tbp}{loa}
\providecommand{\algorithmname}{Algorithm}
\floatname{algorithm}{\protect\algorithmname}

\expandafter\ifx\csname package@font\endcsname\relax\else
 \expandafter\expandafter
 \expandafter\usepackage
 \expandafter\expandafter
 \expandafter{\csname package@font\endcsname}%

\fi
\author[1]{Alex M. Plum}
\author[2]{Christopher P. Kempes}%
\author[3]{Zhen Peng}
\author[3,4]{David A. Baum}

\footnotesize
\affil[1]{\footnotesize Department of Physics, University of California San Diego, La Jolla, CA 92093, USA}
\affil[2]{\footnotesize The Santa Fe Institute, Santa Fe, NM 87501, USA}
\affil[3]{\footnotesize Wisconsin Institute for Discovery, University of Wisconsin-Madison, WI 53706, USA}
\affil[4]{\footnotesize Department of Botany, University of Wisconsin-Madison, WI 53706, USA}

\title{Spatial Structure Supports Diversity in Prebiotic \\ Autocatalytic Chemical Ecosystems}
 
\definecolor{black}{rgb}{0.75,0,0}
\definecolor{blue}{rgb}{0,0,0.75}
\definecolor{green}{rgb}{0,0.5,0}
\definecolor{black}{rgb}{0,0,0}

\begin{document}
\maketitle
\begin{abstract}
\noindent Autocatalysis is thought to have played an important role in the earliest stages of the origin of life. An autocatalytic cycle (AC) is a set of reactions that results in stoichiometric increase in its constituent chemicals. When the reactions of multiple interacting ACs are active in a region of space, they can have interactions analogous to those between species in biological ecosystems. Prior studies of autocatalytic chemical ecosystems (ACEs) have suggested avenues for accumulating complexity, such as ecological succession, as well as obstacles such as competitive exclusion. We extend this ecological framework to investigate the effects of surface adsorption, desorption, and diffusion on ACE ecology. Simulating ACEs as particle-based stochastic reaction-diffusion systems in spatial environments—including open, two-dimensional reaction-diffusion systems and adsorptive mineral surfaces—we demonstrate that spatial structure can enhance ACE diversity by i) permitting otherwise mutually exclusive ACs to coexist and ii) subjecting new AC traits to selection.
\end{abstract}

\section*{Introduction}

In models of prebiotic chemical evolution, spatial structure is frequently invoked to facilitate the accumulation of complexity \cite{kuhn1976model,segre1998graded,vasas2012evolution,walker2012universal,asche2021robotic,asche2024evidence}. For modern cells, spatial structure is endogenous, with autocatalytic metabolisms constructing their own encapsulating membranes. However, life could have started in many ways with endogenous membranes occurring early or late in the evolutionary trajectory and coming before or after complicated autocatalysis. The question is how much evolutionary benefit does spatial separation provide? This includes spatial separation that can be provided  by the environment independent of the prebiotic chemistry. For example, even without endogenous membranes spatial separation could have been supplied by porous rocks, adsorbing mineral surfaces \cite{wachtershauser1988before}, or externally generated vesicles \cite{damer2015coupled}.  Here, we focus on two-dimensional mineral surfaces, which have been the focus of several origin-of-life theories, most notably Wächtershäuser’s surface metabolism model \cite{wachtershauser1988before,wachtershauser2007chemistry}. Mineral-water interfaces would have been ubiquitous in the prebiotic environment and could have played a concentrating role for life-like chemistry, constraining diffusion, catalyzing reactions, and protecting chemicals from hydrolysis and thermal degradation \cite{wachtershauser1988before,wachtershauser1990evolution,wachtershauser1992groundworks,ferris1989mineral}. 
Our aim is not to commit to a particular origin of life nor evolutionary history, but to understand the impact that two-dimensional separation can have on chemical dynamics. The focus on spatial structure is motivated by a large body of work in ecology and evolution illustrating that the addition of space can allow for cooperation, co-existence, and rich dynamics and patterning in cases where the zero-dimensional models do not \cite{turing1952chemical,nowak1993spatial,amarasekare2003competitive,langer2008spatial,roca2009effect,nadell2010emergence,pacala1997biologically}.
\\ \\
Life's earliest autocatalytic chemistry must at least have had a capacity for self-propagation, which requires the existence of at least one autocatalytic cycle (AC): a cyclic reaction pathway that results in a stoichiometric increase of a set of chemicals with each turn of the cycle \cite{blokhuis2020universal}. We call a localized chemical reaction network (CRN) with multiple potential ACs an autocatalytic chemical ecosystem (ACE) \cite{gagrani2024polyhedral,peng2020ecological}. The parallel to ecosystem ecology follows because, like biological species, many ACs can exhibit logistic growth \cite{lloyd1967american}, and pairs of ACs can interact like pairs of biological species, showing competitive, predator-prey, and mutualistic interactions \cite{peng2020ecological}. Moreover, ACEs can undergo ecological succession when the transient introduction of new chemical “seeds” activates new ACs that use the chemicals generated by existing ACs to propagate \cite{peng2022hierarchical}.
Spatial structure can have a profound effect on the dynamics of biological ecosystems. Well-mixed and spatially structured models of ecosystems behave similarly when a single stable fixed point exists, but not necessarily when the system is multi-stable \cite{durrett1994importance}. For example, when there are two stable equilibria in a well-mixed, deterministic model of mutually inhibiting species, the ecological outcome is fully determined by average initial conditions. In spatial models of the same ecosystem, in contrast, the outcome can vary among simulations with the same average initial conditions, due to differences in their spatial pattern \cite{durrett1994importance, levin1974dispersion}. These results from ecosystem ecology show that well-mixed models like chemostats are appropriate for ACEs with just one stable fixed point but that models of multi-stable ACEs, which are of greater relevance to life’s origins, must consider spatial structure. 
\\ \\
As a context to explore the role of spatial structure, we use simple ACEs that consist of mutually inhibiting ACs, designed to exhibit bistability. We show that spatial structure can support the coexistence of otherwise incompatible ACs, increasing ACE diversity. Moreover, we show that spatially structure can affect the fate of competing ACs and result in selection on the basis of spatially relevant traits like diffusivity to which selection is blind in the well-mixed case. These results illustrate how spatial structure may have been important to the early accumulation of complexity, and may have played a role in the emergence of individuated (autopoietic) entities composed of co-dispersing ACs \cite{baum2023ecology}.
\begin{figure}
	\centering
		\includegraphics[width=\textwidth]{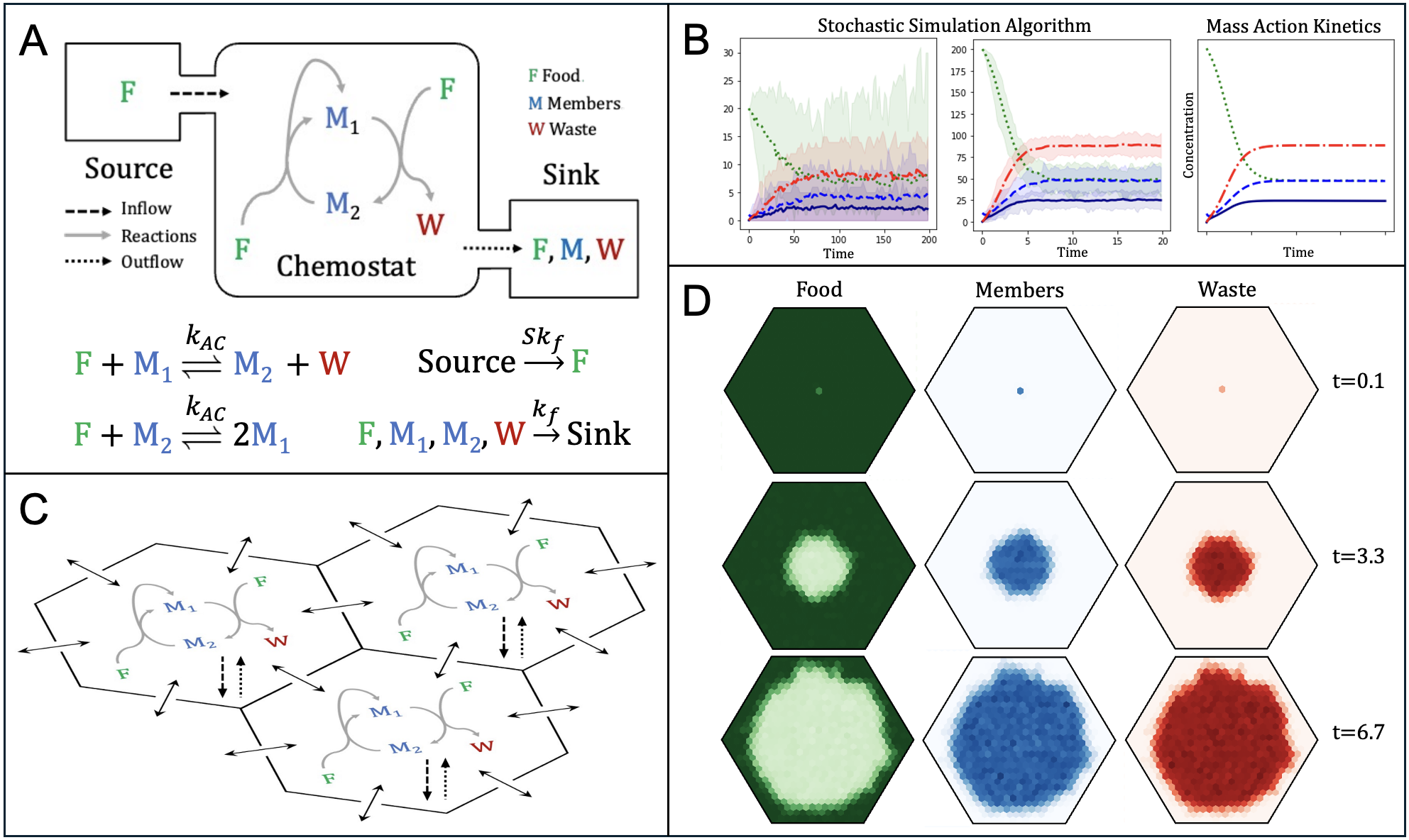}
	\caption{\textbf{A single autocatalytic cycle (AC) in a chemostat and in a 2D array.} \textbf{(A)} An AC with one food chemical ($F$), two member chemicals ($M_1$, $M_2$), and one waste chemical ($W$). Cycle reactions are reversible with rate constant $k_{AC}$ in each direction. A flux of food through a chemostat drives AC reactions in the forward direction, indicated by gray arrows. Food flows in from a source with constant food count $S$ and all chemical flow out to a sink with flow rate $k_f$. \textbf{(B)} SSA simulations of the AC in \textbf{(A)}. Discrete chemical counts converge to mass action kinetics concentration curves with increasing $S$. Curves are mean counts over 10 replicate SSA simulations. Shaded regions show the range of values observed. Left: SSA using source food count $S=20$, seeded with 1 $M_1$. Middle: SSA with $S=200 \ F$, seeded with 10 $M_1$. Right: Mass action kinetics with $S=200$ units $F$, seeded with 10 units $M_1$. Concentration units are arbitrary. \textbf{(C, D)} Spatial growth in a 2D array. \textbf{(C)} A section of a hexagonal reaction-diffusion array, treating each site as a chemostat (as in \textbf{(A)}) with diffusion events between them. \textbf{(D)} An AC seeded in the center of an array of diameter 31 with periodic boundaries. Heatmaps show the counts of food ($F$, left, green), total member chemical ($M_1+M_2$, middle, blue), and waste ($W$, right, red) at three time points (darker means more). Each site begins with the source food count ($S=1000$) and the central cite contains a seed of $250 \ M_1$ molecules. Each simulation was run for a time $T=6.7$ with $k_{AC}=k_f=0.01$, a diffusion rate constant $k_D=0.6$ for all molecules, and $\tau$-leaping.}
	\label{FIG:1}
\end{figure}

\section*{Methods}

\subsection*{Autocatalytic Cycles \& Autocatalytic Chemical Ecosystems}

We adopt a stoichiometric definition of autocatalysis \cite{blokhuis2020universal} that is compatible with reversible chemical kinetics \cite{peng2020ecological}. We categorize an AC's chemicals as member chemicals if they appear as both reactants and products in AC reactions. Food and waste are categorized relative to the reaction directions that results in a stoichiometric increase in member chemicals, deemed the AC's forward direction: food chemicals only appear as reactants in the forward direction, whereas waste chemicals only appear as products (Fig. \ref{FIG:1}A). When all of the AC's reactions can proceed in the forward direction, an AC can self-propagate and we deem it \textit{active}. When an adequate supply of food and removal of waste drives an AC's reactions in the forward direction, for example in a chemostat (Fig. \ref{FIG:1}A), the AC self-propagates. An active AC’s ability to self-propagate is independent of the types of chemicals involved, which can range from small molecules to long, replicating polymers or multi-molecular assemblies.

\subsection*{Discreteness \& Stochasticity}

Prior work simulated ACE dynamics in chemostats using mass-action kinetics \cite{peng2020ecological}. Here, we opt instead for stochastic methods and discrete chemical counts. Stochastic methods capture environmental and demographic stochasticity characteristic of real biological ecosystems and chemical reactions, which can affect ecological outcomes and pattern formation \cite{butler2009robust,butler2011fluctuation,jafarpour2015noise,jafarpour2017noise}. Likewise, they are more relevant to origins-of-life scenarios where rare reactions and small numbers of chemicals could have played significant roles \cite{wu2009origin, wu2012origin}. With stochasticity, an active AC can deactivate after losing its last member chemical (e.g. Fig. \ref{FIG:1}B, left). This contrasts with mass action models where the chemical concentrations of member chemicals never drop to zero but, instead approach it asymptotically. Likewise, a single dispersed seed molecule can colonize a new location and activate a previously inactive AC (Fig. \ref{FIG:1}D). 
\\ \\
The most common stochastic method for simulating chemical reaction networks is Gillespie’s exact Stochastic Simulation Algorithm (SSA) \cite{gillespie1976general}. At low concentrations (smaller counts at a fixed volume) stochasticity has a great effect that diminishes with higher molecule counts. As (Fig. \ref{FIG:1}A) illustrates for a simple AC in a chemostat, the SSA’s discrete count trajectories converge to an ODE's continuous concentration curves as the fixed number of food molecules available to flow in from the source ($S$) increases (Fig. \ref{FIG:1}B).

\subsection*{Spatial Structure}

Here, we model a particle-based stochastic reaction-diffusion system, where stochastic reactions occur locally in a hexagonal array of discrete sites and stochastic diffusion transfers chemicals between sites. The combination of discreteness, stochasticity, and spatial structure in particle-based stochastic reaction-diffusion systems allows for richer and more realistic ecological dynamics. Each site in the hexagonal array is treated as a chemostat and simulated with the SSA. All chemicals diffuse stochastically between adjacent sites and inflow, outflow, and reaction events occur within each site (Fig. \ref{FIG:1}C) \cite{erban2007practical}. To demonstrate the model, a viable AC seeded at one location expands radially as its member chemicals stochastically colonize adjacent sites (Fig. \ref{FIG:1}D). For our other spatial simulations with larger CRNs or spatial environments, we employ $\tau$-leaping, an efficient approximation of the SSA that updates reaction propensities less frequently \cite{anderson2008incorporating, cao2006efficient}. 

\subsection*{Mutually inhibiting Autocatalytic Cycles}
\begin{figure}
	\centering
		\includegraphics[width=\textwidth]{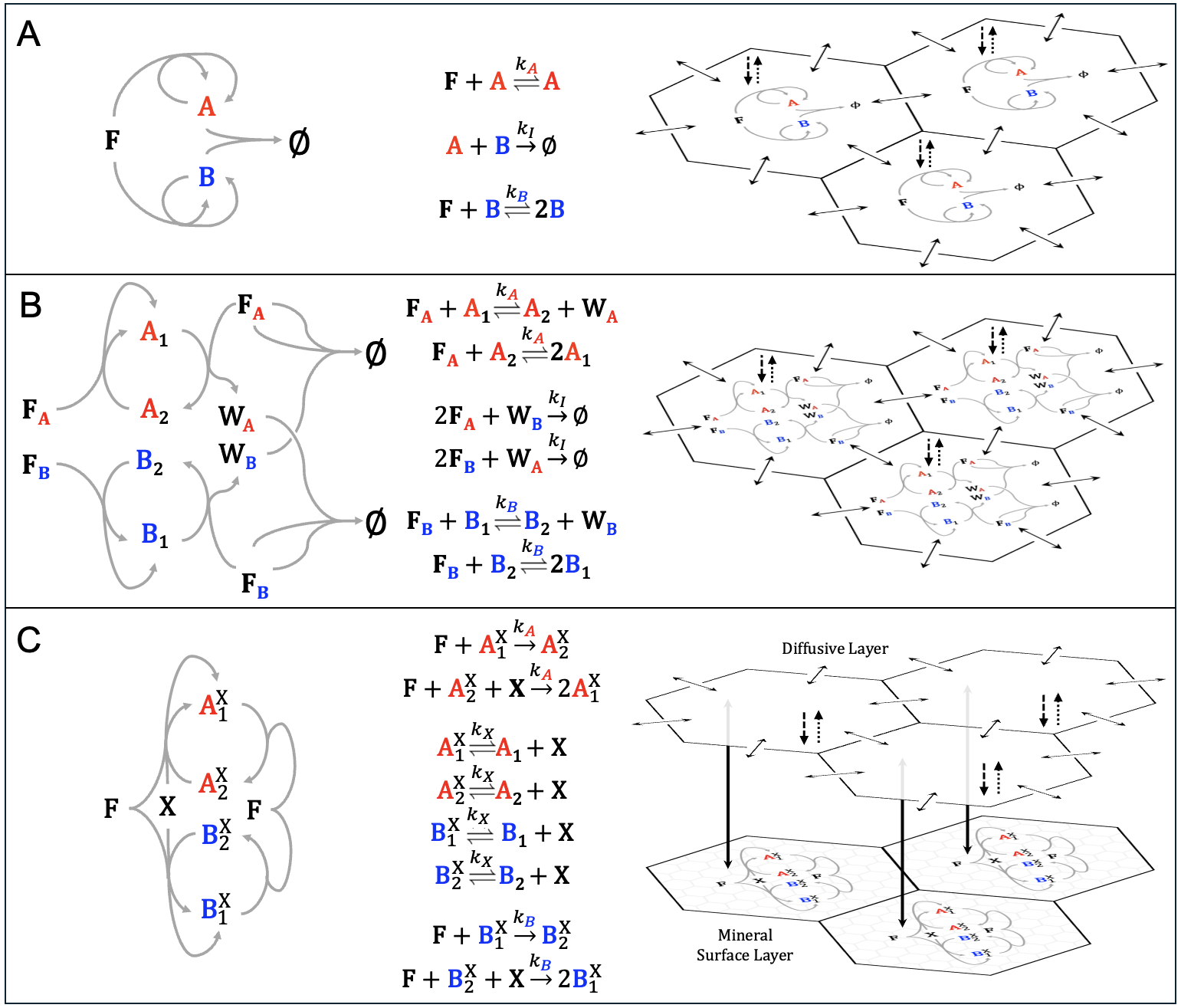}
	\caption{\textbf{Three example ACEs exhibiting bistability.}
    Pairs of ACs might exhibit bistability via \textbf{(A)} member-member reactions, \textbf{(B)} waste-food reactions, or \textbf{(C}) surface competition. Left: Chemical Reaction Networks. Arrows indicate inhibition reactions and the forward direction of AC reactions. Middle: Chemical reactions and their rate constants, in addition to food inflow and outflow. Right: Subsets of spatial environments. In each site, food flows in from a source (dashed arrows), all chemicals flow out to a sink (dotted arrows), react (gray arrows), and diffuse to adjacent sites (solid black arrows). In the surface competition model \textbf{(C)} there are two-layers: a diffusive layer with diffusion, inflow, and outflow, and a surface layer with surface-catalyzed AC reactions and a finite number of adsorption sites $X$. Chemicals move between the layers via adsorption and desorption reactions.}
	\label{FIG:2}
\end{figure}

We designed several ACEs exhibiting bistability, each consisting of a pair of mutually inhibiting ACs, $A$ and $B$. Any ACE with a viable AC is trivially multi-stable because that AC could be active or inactive. Here, we use bistability to refer to cases in which each AC could remain active on its own, but cannot coexist with the other in a well mixed setting, providing just two locally stable outcomes: either $A$ or $B$  deactivates.
\\ \\
Our first mechanism for mutual inhibition (Fig. \ref{FIG:2}A) uses ACs whose member chemicals dimerize into an unreactive product. Because food and waste are not involved in this inhibition mechanism, we opted to simplify the system to a pair of single-member ACs with shared food and no waste. Although these ACs share food, we know from previous work that, with reversible reactions, simple competition for food does not lead to bistability in ACEs \cite{peng2020ecological}. This mechanism for bistability---reminiscent of simple models for the emergence of homochirality \cite{frank1953spontaneous,jafarpour2015noise}---contains the fewest number of reactions among the three mechanisms considered, making its simulations the least computationally demanding.
\\ \\
A second mechanism for mutual inhibition (Fig. \ref{FIG:2}B) was previously explored in \cite{peng2020ecological} for its ability to demonstrate ecological priority effects. For this mechanism, we use two-member ACs (ACs that generate waste while retaining second order reactions) that do not share food or waste. Mutual inhibition arises because each AC’s waste reacts with two of the other AC’s food to produce another chemical (not explicitly modeled) that neither AC can directly use. As one AC grows, its accumulating waste saps the other’s food, thereby lowering the other's growth rate. 
\\ \\
A third mechanism for mutual inhibition arises from mutual dependence on a catalytic surface. We separate the reaction diffusion system into a diffusive layer subject to inflow, outflow, and diffusion and a surface layer, where AC reactions take place but there is no diffusion (Fig. \ref{FIG:2}C). The member chemicals of each AC adsorb and desorb between the diffusive and surface layers (Fig. \ref{FIG:2}C). A finite number of adsorption sites are defined at each location in the surface layer. We neglect reactions in the diffusive layer and treat surface reactions as irreversible. Mutual inhibition arises because of zero-sum competition for these surface sites, which, unlike food, are not continuously resupplied by the external environment. Each AC’s self-propagation requires that a member chemical be adsorbed and an open adsorption site for the new member chemical produced. An AC that occupies open adsorption sites can, therefore, inhibit the other AC. 

\section*{Results}

\subsection*{Spatial coexistence of mutually inhibiting Autocatalytic Cycles}
First, we explore the effect of chemical diffusivity on ACE dynamics by simulating a reaction-diffusion domain that is large enough that spatial patterns can be seen and analyzed statistically. We use the first bistability mechanism (member-member reactions, Fig. \ref{FIG:2}A). We simulate open reaction-diffusion arrays seeded uniformly with members of Cycles $A$ and $B$, and vary the diffusion rate constant $k_D$ of these member chemicals. To highlight the effects of diffusion, we set all reaction rate constants equal.
\\ \\
To quantify spatial patterns, we use several descriptive metrics. With member counts $A_i$ and $B_i$ at site $i\in[1,...N]$, we consider the local member fractions $P_{A,i}=\frac{A_i}{A_i+B_i}$ and $P_{B,i}=\frac{B_i}{A_i+B_i}$ and local occupancy $O_i=P_{A,i}-P_{B,i}=\frac{A_i-B_i}{A_i+B_i}$.  Occupancy values for a site range from -1 (full occupation by $B$) to 1 (full occupation by $A$). We then consider three measures of diversity. First, we consider the Shannon index of local diversity $H_{loc,i}=-P_{A,i}\log_2{P_{A,i}}-P_{B,i} \log_2 {P_{B,i}}$. Second, to capture the degree of correlation between neighboring sites, we calculate a neighborhood heterogeneity $H_{nei,i}=\frac{1}{12}\sum_{j=1}^6 |O_i-O_j|$, where $j$ ranges over the six neighbors around site $i$. Third, we calculate the global diversity: $H_{glob}=-P_A \log_2 {P_A}-P_B \log_2{P_B}$ where $P_{A}=(\sum_{i=1}^N A_i)/(\sum_{i=1}^N A_i+B_i)$ and $P_{B}=(\sum_{i=1}^N B_i)/(\sum_{i=1}^N A_i+B_i)$. All three measures range from 0 and 1. In ecological terms, $H_{loc}$ and $H_{glob}$ represent alpha and gamma diversity \cite{daly2018ecological}. 
\begin{figure}[!ht]
	\centering
	\includegraphics[width=\textwidth]{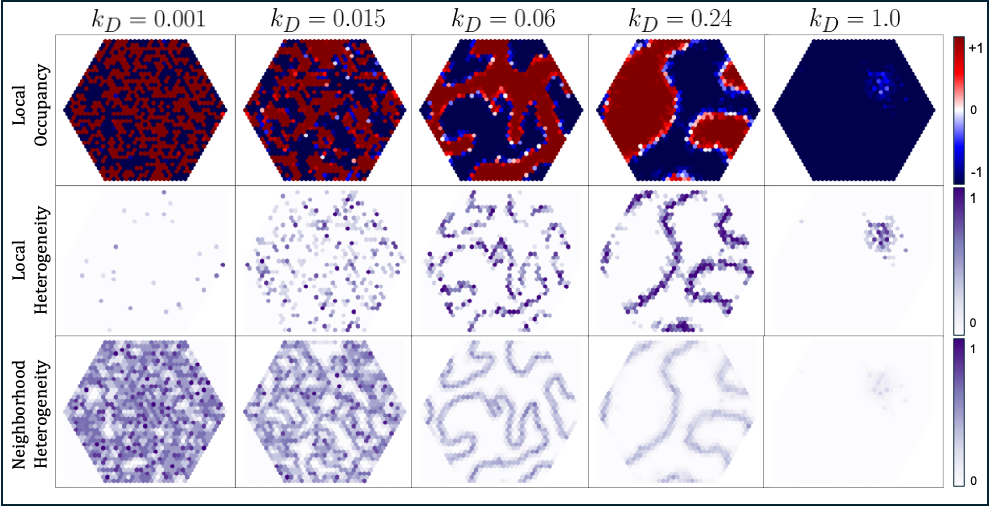}
    	\caption{\textbf{Final state of ACEs simulated with different diffusivities.}
        Heatmaps of local occupancy ($O_i$), local diversity ($H_{loc,i}$), and neighborhood heterogeneity ($H_{nei,i}$) in an open reaction-diffusion array of diameter 39 as member diffusivity ($k_D$) varies from $10^{-3}$ to 1. Each site begins with 10 $A$, 10 $B$, and $50 \ F$. Because food continuously flows in, we neglect its diffusion. Each simulation was run to time $T = 100$ with $k_{AC,A}=k_{AC,B}=0.01$,$k_f=0.01$, $k_I=0.02$, $S=50$, and $\tau$-leaping.}    
	\label{FIG:3}
\end{figure} 
\\ \\
Fig. \ref{FIG:3} shows the spatial distribution of $O_i$, $H_{loc,i}$, and $H_{nei,i}$ at the final time step for six diffusion rates.  The slow diffusion regime reaches low $H_{loc}$ and high $H_{glob}$ because the ACs do not coexist locally but remain active in nearly equal quantities globally. In contrast, the high diffusion regime converges towards low $H_{loc}$ and low $H_{glob}$ because one AC drives the other extinct everywhere, though it does so slowly such that $H_{loc}$ remains transiently high. The intermediate diffusion regime attains intermediate values of $H_{glob}$ and $H_{nei}$. Notably, $H_{loc}$ peaks for intermediate diffusion.
\\ \\
With low diffusion, one AC quickly occupies and deactivates the other AC in each site. Because so few chemicals are exchanged between sites, neither AC can successfully invade a site occupied by the other, so each site quickly converges to a steady state with low $H_{loc}$. However, because sites can become stably occupied by either AC, $H_{glob}$ remains high. With high diffusion, adjacent sites exchange large numbers of chemicals and exhibit spatially autocorrelated dynamics. In the extreme case, the dynamics approximate a single well-mixed chemostat, resulting in global bistability with $H_{glob}$  dropping to 0. In intermediate diffusion regimes, however, sites may transition between periods of occupation by each AC as they successfully invade one another’s sites. This results in patches of the surface occupied by each AC. The patch sizes to which simulations converge increase with $k_D$ (Fig. \ref{FIG:3}), reminiscent of the temperature dependent correlation length of the 2D Ising model \cite{onsager1944crystal}. With both  low and intermediate diffusion, the ACE exhibits long-term coexistence, but the intermediate regime entails more frequent reacitons between the ACs’ chemicals resulting in a greater chemical diversity (i.e. dimer production) near the boundaries between patches. 

\subsection*{Faster vs. Fiercer: Selection for diffusivity}
\begin{figure}
    \centering
    \includegraphics[width=\textwidth]{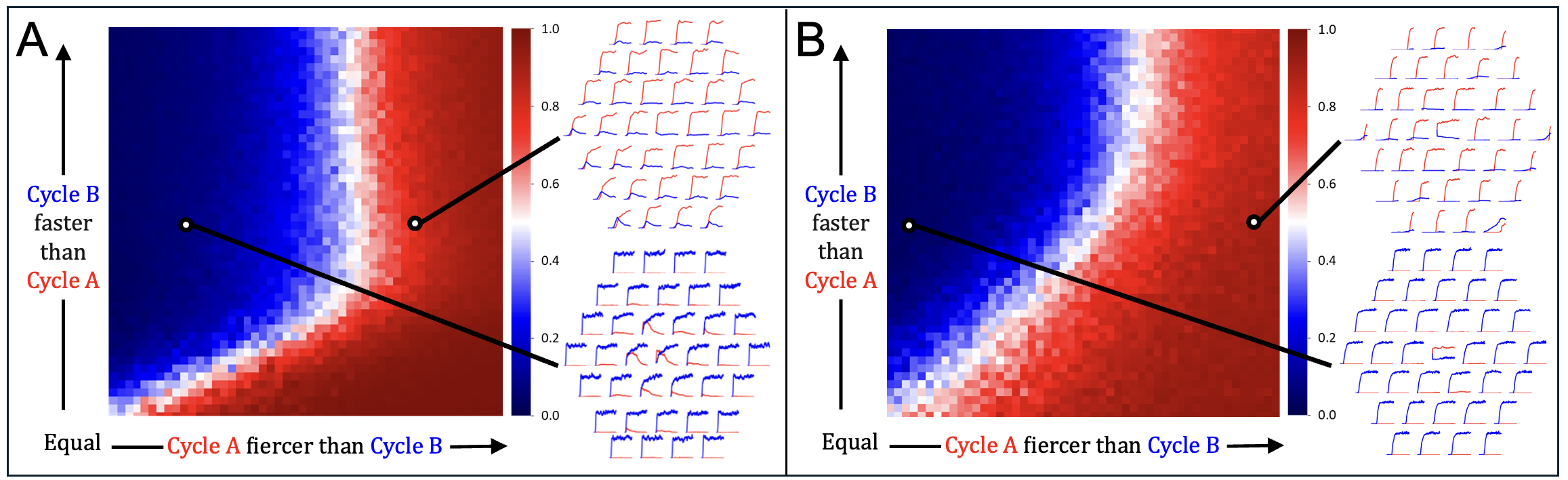}
    \caption{\textbf{Faster vs. fiercer} \textbf{(A)} Food-waste inhibition (\ref{FIG:2}B), \textbf{(B)} Surface inhibition (\ref{FIG:2}C). Results of simulations in an open reaction-diffusion of diameter 7 with periodic boundaries. \textbf{(A, B)} Left: Heat maps of the average, relative abundances of each AC’s member chemicals at the end of the simulation for different parameter combinations, averaged over 10 replicates. Right: Example time series of the member species of each AC at each site.
    Across all simulations, $A$ has fixed reaction rate constant $10^{-2}$ and fixed diffusion rate constant $10^{-1}$, and $B$’s diffusion rate constant varies from $10^{-1}$ to $10^3$. All simulations use $k_f = 10^{-1}$ and $\tau$-leaping and are initialized with source food and a seed of $10 \ A_1$ and $10 \ B_1$ in the central site. \textbf{(A)} $B$’s reaction rate constant varies from $10^{-2.5}$ to $10^{-2}$. Top: $A$ dominates for parameter combination ($0.004$, $10$). Bottom: $B$ dominates for parameter combination ($0.008$, $10$). Simulations run with inhibition rate $k_I=1$ and $S=1000$ for total time $T = 30$. \textbf{(B)} $B$’s reaction rate constants varies from $10^{-3.5}$ to $10^{-2}$.  The central site was seeded with $10 A_1$ and $10 B_1$. Top: $A$ dominates for parameter combination ($0.0004$, $10$). Bottom: $B$ dominates for parameter combination ($0.008$, $10$). Simulations run with adsorption and desorption rate $k_X = 10^{-2}$, $100$ adsorption sites, and $S=100$ for total time  $T=10$.}   
\label{FIG:4}
\end{figure}

Next, we investigate the role of member different chemical diffusivities and reaction rates in ecological interactions among mutually inhibiting ACs. In contrast to the prior simulations, where both ACs had the same reaction and diffusion rate constants, here we vary them independently. We say that an AC is \textit{fiercer} if its reaction rate constants are higher. We say that an AC is \textit{faster} if its member chemicals have higher diffusivity. We explored a range of reaction and diffusion rates in the regime in which $A$ is fiercer than $B$ and $B$ is faster than $A$. In this regime, $A$ tends to self-propagate in a local site more rapidly using available food whereas $B$ tends to reach new sites faster, which allows it to exploit new sources of food. 
\\ \\
First, we use the second mechanism for mutual inhibition: waste-food reactions (Fig. \ref{FIG:2}B). We simulate these ACs in open reaction-diffusion systems of diameter 7 with periodic boundaries, varying the reaction-rate advantage of $A$ and the diffusion-rate advantage of $B$. For each parameter combination, we seed both cycles in equal quantities in the central site, initializing food uniformly. Despite the fact that $A$, being fiercer, always tends to deactivate $B$ in the central site, $A$ can ``escape'' by spreading out and occupying other sites that have abundant food and are not yet occupied by $A$ (Fig. \ref{FIG:4}A). When this happens $B$ can establish occupancy in all surrounding sites and can sometimes re-invade the central site to deactivate $A$ globally (Fig. \ref{FIG:4}A, bottom right). Looking across a range of values we see that $B$ can only achieve a finite occupancy advantage and that beyond some threshold of relative fierceness $A$ will always dominate. Below this threshold, the faster-diffusing AC can out-compete the faster-reacting one. 
\\ \\
We repeat these experiments with the third mechanism for mutual inhibition: adsorption site competition (Fig. \ref{FIG:2}C) and obtain similar results (Fig. \ref{FIG:4}B). In a chemostat, the fiercer AC will always be favored. Nevertheless, as with the first inhibition mechanism, we find that there is a parameter range in which higher diffusivity is favored over higher reactivity (Fig. \ref{FIG:4}B). 

\section*{Discussion}

\subsection*{Spatial structure enriches ACE dynamics}
Our \textit{in silico} experiments show that spatial structure enriches ACE dynamics in several ways. Most obviously, spatial structure increases global chemical diversity and can eliminate convergence to a single global attractor. In intermediate diffusion regimes, spatial structure can also accommodate transient patches and dynamic patterns. Chemical diversity consists of both the member and waste chemicals of globally coexisting ACs, as well as any products of interactions between them that cannot be generated by each individual AC. The latter are produced only when there is local coexistence of at least two ACs. In a well-mixed setting, such chemical diversity tends to vanish because one of the necessary ACs is deactivated globally. Likewise, in sets of isolated reactors (no diffusion), one or other of the two ACs will be deactivated locally. Thus, it is only in intermediate diffusion regimes that both ACs would continue to interact with one another and thereby contribute to the overall chemical diversity. This parallels the more general observation in biology that ``\textit{Complex systems when combined with slight migration between them produce even more complex systems}'' \cite{karlin1972application}. Mineral surfaces provide one way of achieving intermediate diffusion rates by lowering effective diffusion rates through chemical adsorption. 
\\ \\
A second implication of spatial structure is that it adds new factors that might undergo selection. When we vary diffusion symmetrically, the result resembles pure genetic drift in that neither AC has an intrinsic advantage over the other, yet one or the other is expected to go to fixation in the long run. In contrast, when we consider ACs that differ in reaction and diffusion rates, the result resembles selection, since an AC's ability to persist and propagate is influenced by intrinsic (``fitness'') differences. This demonstrates that diffusivity can be a target of selection in settings where chemical or catalytic resources remain untapped. Moreover, these results suggest that with patchily distributed resources or intermittent disturbance, ACEs could become dominated by fast dispersing but kinetically ``weak'' ACs rather than the fierce ACs that tend to dominate under low disturbance.  This finding suggests a possible chemical analog of the intermediate disturbance hypothesis \cite{connell1978diversity}.  
\\ \\
Spatial structure has been shown elsewhere to create new ACs when it affects the underlying CRN \cite{blokhuis2020universal}. Blockhuis et al. provide an example of a CRN affording no ACs in a well-mixed setting that nevertheless has one AC when operating in two connected compartments that are each conducive to a different subset of reactions. A mineral surface can effectively create two compartments by creating a distinction between adsorbed and desorbed chemicals, permitting catalyzed reactions in the adsorbed layer, and shielding adsorbed chemicals from direct inflow or outflow. Similarly, one can imagine additional dynamical and selective complexity emerging in other spatial structures, such as selectively permeable compartments.

\subsection*{ACEs in spaces blur the boundaries between ecology
and evolution} 

Biological ecosystems, like organisms themselves, are complex adaptive systems \cite{levin1998ecosystems} whose accumulation of complexity does not depend solely on the Darwinian evolution of the populations that compose them. Even in the absence of Darwinian evolution, ecosystem complexity can arise through succession \cite{gleason1927further}, self-organization \cite{levin2005self}, endogenous pattern formation \cite{levin1985pattern}, and autocatalytic niche emergence \cite{cazzolla2020emergence}, where niche emergence drives ecological diversity and that diversity drives the emergence of new niches \cite{gatti2018niche}. By analogy, ACEs can accumulate complexity in a manner analogous to biological ecosystems without the evolvability of ACs themselves \cite{peng2020ecological,peng2022hierarchical}. The untapped creative potential of early ACs lies not in their ability to evolve, but in their potential to stably organize in innumerable combinations and contexts, each with their own constructive and catalytic capacities.
\\ \\
Natural selection can act on chemical processes insofar as they have a capacity to self-propagate, change, and propagate their changes. While individual ACs can self-propagate, they lack capacity for variation, making individual ACs poor candidates for units of selection \cite{godfrey2009darwinian}. ACEs have a much greater capacity for variation because they can contain different combinations of active ACs that confer different ecosystem-level properties \cite{baum2023ecology}. As a result, the set of possible, propagable changes to any ACE will usually be much larger than for any individual AC. ACEs’ capacity for variation makes the more plausible units of selection. This resonates with the ecosystem-first theory of life’s origins \cite{hunding2006compositional} in which the primitive adaptive dynamics of interacting chemical processes preceded and provided for the emergence of paradigmatic Darwinian evolution. When there are many separated, locally stable ACEs that interact and exchange material, as discussed by \cite{baum2023ecology}, they can be seen to comprise a primitive Darwinian population, an autocatalytic chemical meta-ecosystem (ACME). 
The degree of separation and stable boundaries that ACEs maintain has implications for their evolvability. Trivially, spatial structure is necessary for boundaries to exist between ACEs at all and, as we have shown, chemical diffusivities can affect the extent and stability of those boundaries. When ACEs form patches with stable boundaries, they may behave like discrete units even without cellular encapsulation \cite{baum2015selection}. Considering ACMEs as analogs of Darwinian populations allows us to link the theoretical frameworks of spatial ecosystem ecology with the prebiotic emergence of adaptive evolution.

\subsection*{Avenues for Future Work}

Future work on ACEs in spatially structured settings can expand both the CRNs considered and the spatial structures in which they are situated. Here we have confined our experiments to pairs of mutually inhibiting ACs. Even within this narrow scope, other mechanisms for mutual inhibition are possible. For example, one AC’s waste can react with another AC’s members. Additionally, there are other routes to nontrivial bistability that do not involve mutual inhibition, such as the Schlögl model \cite{schlogl1972chemical}, which achieves bistability with just one AC. Likewise, predation, which has been demonstrated in ACEs \cite{peng2020ecological}, is known to be capable of exhibiting complex spatial patterns \cite{murray2001mathematical}. When more than two ACs are present, there is also the possibility of multistability via unidirectional inhibition, for example, when three ACs inhibit one other cyclically, coexisting in an ecological game of rock-paper-scissors \cite{kerr2002local}. Other ACEs of interest may emulate Turing mechanisms \cite{turing1952chemical, munuzuri2022unified}. Finally, a great deal of origins of life literature has focused on life's homochirality, a lack of coexistence between chiral counterparts \cite{frank1953spontaneous,gleiser2008extended,gleiser2012life}. In this context, stochastic dynamics in spatial systems can eliminate coexistence \cite{jafarpour2015noise,jafarpour2017noise}, stabilizing a homochiral state even when well-mixed deterministic models would favor a racemic state. While stochasticity widens the space of ecological possibilities, it will be important to elucidate and characterize both its creative and destructive effects on chemical diversity.
\\ \\
Here, we have focused on 2D reaction-diffusion systems and mineral surfaces, assuming spatially uniform inflow and outflow. However, localized sources and sinks are also possible, which could yield patterns resembling ecological clines. For example, ACMEs with directional flow between ACEs, and receiving input of simple precursors at one end, could establish a gradient from low to high chemical complexity as one moves downstream to more distant ACEs. 
Another obvious class of models to explore would be protocell-like spatial structures, in which some chemicals are only periodically (if ever) exchanged between reactors. There are reasons to suppose that more discrete spatial structures may provide greater protection for the prebiotic chemistry they harbor \cite{shah2019survival, takeuchi2009multilevel} and result in ACMEs that resemble populations of protocells, facilitating natural selection. The simplest models would assume that these discrete structures are provided by the external environment, but more sophisticated models might allow for the possibility that ACEs create or reinforce their own boundaries, for example by producing vesicle-forming amphiphiles. Such spatial models might ultimately allow us to build a plausible pathway from systems whose spatial structure is imposed externally via mineral surfaces or externally generated compartments, to ACEs more like modern cells, which individuate themselves. Indeed, we would argue that a worthwhile long-term goal of studies of the spatial aspects of prebiotic chemical dynamics should be to explain the emergence of higher-level units of selection \cite{luisi1989self} in the transition from exogenously to endogenously controlled spatial structure.

\section*{Acknowledgements}
We thank Praful Gagrani, Stephanie Col$\text{\'o}$n-Santos, Lena Vincent, and Nigel Goldenfeld for helpful discussions about prebiotic chemistry, chemical ecology, and non-equilibrium pattern formation. We also thank all members of the Baum Lab, the Santa Fe Institute Evolving Chemical Systems working group, the 2020 Santa Fe Institute Undergraduate Complexity Research Program, and NASA grant No. 80NSSC17K0296. David Baum acknowledges support from NSF grant 2218817. Computation was made possible by the University of Wisconsin-Madison Center for High-Throughput computing \cite{https://doi.org/10.21231/gnt1-hw21}. 
 
\clearpage
\bibliographystyle{ieeetr}
\bibliography{references}

\providecommand{\noopsort}[1]{}\providecommand{\singleletter}[1]{#1}%
\begin{thebibliography}{10}

\bibitem{kuhn1976model}
H.~Kuhn, ``Model consideration for the origin of life: Environmental structure as stimulus for the evolution of chemical systems,'' {\em Naturwissenschaften}, vol.~63, pp.~68--80, 1976.

\bibitem{segre1998graded}
D.~Segr{\'e}, D.~Lancet*, O.~Kedem, and Y.~Pilpel, ``Graded autocatalysis replication domain (gard): kinetic analysis of self-replication in mutually catalytic sets,'' {\em Origins of Life and Evolution of the Biosphere}, vol.~28, pp.~501--514, 1998.

\bibitem{vasas2012evolution}
V.~Vasas, C.~Fernando, M.~Santos, S.~Kauffman, and E.~Szathm{\'a}ry, ``Evolution before genes,'' {\em Biology direct}, vol.~7, pp.~1--14, 2012.

\bibitem{walker2012universal}
S.~I. Walker, M.~A. Grover, and N.~V. Hud, ``Universal sequence replication, reversible polymerization and early functional biopolymers: a model for the initiation of prebiotic sequence evolution,'' {\em PloS one}, vol.~7, no.~4, p.~e34166, 2012.

\bibitem{asche2021robotic}
S.~Asche, G.~J. Cooper, G.~Keenan, C.~Mathis, and L.~Cronin, ``A robotic prebiotic chemist probes long term reactions of complexifying mixtures,'' {\em Nature Communications}, vol.~12, no.~1, p.~3547, 2021.

\bibitem{asche2024evidence}
S.~Asche, R.~W. Pow, H.~M. Mehr, G.~J. Cooper, A.~Sharma, and L.~Cronin, ``Evidence of selection in mineral mediated polymerization reactions executed in a robotic chemputer system,'' {\em ChemSystemsChem}, vol.~6, no.~3, p.~e202400006, 2024.

\bibitem{wachtershauser1988before}
G.~W{\"a}chtersh{\"a}user, ``Before enzymes and templates: theory of surface metabolism,'' {\em Microbiological reviews}, vol.~52, no.~4, pp.~452--484, 1988.

\bibitem{damer2015coupled}
B.~Damer and D.~Deamer, ``Coupled phases and combinatorial selection in fluctuating hydrothermal pools: A scenario to guide experimental approaches to the origin of cellular life,'' {\em Life}, vol.~5, no.~1, pp.~872--887, 2015.

\bibitem{wachtershauser2007chemistry}
G.~W{\"a}chtersh{\"a}user, ``On the chemistry and evolution of the pioneer organism,'' {\em Chemistry \& Biodiversity}, vol.~4, no.~4, pp.~584--602, 2007.

\bibitem{wachtershauser1990evolution}
G.~W{\"a}chtersh{\"a}user, ``Evolution of the first metabolic cycles.,'' {\em Proceedings of the National Academy of Sciences}, vol.~87, no.~1, pp.~200--204, 1990.

\bibitem{wachtershauser1992groundworks}
G.~W{\"a}chtersh{\"a}user, ``Groundworks for an evolutionary biochemistry: the iron-sulphur world,'' {\em Progress in biophysics and molecular biology}, vol.~58, no.~2, pp.~85--201, 1992.

\bibitem{ferris1989mineral}
J.~P. Ferris, G.~Ertem, and V.~Agarwal, ``Mineral catalysis of the formation of dimers of 5'-amp in aqueous solution: the possible role of montmorillonite clays in the prebiotic synthesis of rna,'' {\em Origins of Life and Evolution of the Biosphere}, vol.~19, no.~2, pp.~165--178, 1989.

\bibitem{turing1952chemical}
A.~M. Turing, ``The chemical basis of morphogenesis,'' {\em Philosophical Transactions of the Royal Society of London. Series B, Biological Sciences}, vol.~237, no.~641, pp.~37--72, 1952.

\bibitem{nowak1993spatial}
M.~A. Nowak and R.~M. May, ``The spatial dilemmas of evolution,'' {\em International Journal of bifurcation and chaos}, vol.~3, no.~01, pp.~35--78, 1993.

\bibitem{amarasekare2003competitive}
P.~Amarasekare, ``Competitive coexistence in spatially structured environments: a synthesis,'' {\em Ecology letters}, vol.~6, no.~12, pp.~1109--1122, 2003.

\bibitem{langer2008spatial}
P.~Langer, M.~A. Nowak, and C.~Hauert, ``Spatial invasion of cooperation,'' {\em Journal of theoretical biology}, vol.~250, no.~4, pp.~634--641, 2008.

\bibitem{roca2009effect}
C.~P. Roca, J.~A. Cuesta, and A.~S{\'a}nchez, ``Effect of spatial structure on the evolution of cooperation,'' {\em Physical Review E—Statistical, Nonlinear, and Soft Matter Physics}, vol.~80, no.~4, p.~046106, 2009.

\bibitem{nadell2010emergence}
C.~D. Nadell, K.~R. Foster, and J.~B. Xavier, ``Emergence of spatial structure in cell groups and the evolution of cooperation,'' {\em PLoS computational biology}, vol.~6, no.~3, p.~e1000716, 2010.

\bibitem{pacala1997biologically}
S.~W. Pacala and S.~A. Levin, ``Biologically generated spatial pattern and the coexistence of competing species,'' {\em Spatial ecology: the role of space in population dynamics and interspecific interactions}, pp.~204--232, 1997.

\bibitem{blokhuis2020universal}
A.~Blokhuis, D.~Lacoste, and P.~Nghe, ``Universal motifs and the diversity of autocatalytic systems,'' {\em Proceedings of the National Academy of Sciences}, vol.~117, no.~41, pp.~25230--25236, 2020.

\bibitem{gagrani2024polyhedral}
P.~Gagrani, V.~Blanco, E.~Smith, and D.~Baum, ``Polyhedral geometry and combinatorics of an autocatalytic ecosystem,'' {\em Journal of Mathematical Chemistry}, pp.~1--67, 2024.

\bibitem{peng2020ecological}
Z.~Peng, A.~M. Plum, P.~Gagrani, and D.~A. Baum, ``An ecological framework for the analysis of prebiotic chemical reaction networks,'' {\em Journal of theoretical biology}, vol.~507, p.~110451, 2020.

\bibitem{lloyd1967american}
P.~Lloyd, ``American, german and british antecedents to pearl and reed's logistic curve,'' {\em Population Studies}, vol.~21, no.~2, pp.~99--108, 1967.

\bibitem{peng2022hierarchical}
Z.~Peng, J.~Linderoth, and D.~A. Baum, ``The hierarchical organization of autocatalytic reaction networks and its relevance to the origin of life,'' {\em PLOS Computational Biology}, vol.~18, no.~9, p.~e1010498, 2022.

\bibitem{durrett1994importance}
R.~Durrett and S.~Levin, ``The importance of being discrete (and spatial),'' {\em Theoretical population biology}, vol.~46, no.~3, pp.~363--394, 1994.

\bibitem{levin1974dispersion}
S.~A. Levin, ``Dispersion and population interactions,'' {\em The American Naturalist}, vol.~108, no.~960, pp.~207--228, 1974.

\bibitem{baum2023ecology}
D.~A. Baum, Z.~Peng, E.~Dolson, E.~Smith, A.~M. Plum, and P.~Gagrani, ``The ecology--evolution continuum and the origin of life,'' {\em Journal of the Royal Society Interface}, vol.~20, no.~208, p.~20230346, 2023.

\bibitem{butler2009robust}
T.~Butler and N.~Goldenfeld, ``Robust ecological pattern formation induced by demographic noise,'' {\em Physical Review E—Statistical, Nonlinear, and Soft Matter Physics}, vol.~80, no.~3, p.~030902, 2009.

\bibitem{butler2011fluctuation}
T.~Butler and N.~Goldenfeld, ``Fluctuation-driven turing patterns,'' {\em Physical Review E—Statistical, Nonlinear, and Soft Matter Physics}, vol.~84, no.~1, p.~011112, 2011.

\bibitem{jafarpour2015noise}
F.~Jafarpour, T.~Biancalani, and N.~Goldenfeld, ``Noise-induced mechanism for biological homochirality of early life self-replicators,'' {\em Physical review letters}, vol.~115, no.~15, p.~158101, 2015.

\bibitem{jafarpour2017noise}
F.~Jafarpour, T.~Biancalani, and N.~Goldenfeld, ``Noise-induced symmetry breaking far from equilibrium and the emergence of biological homochirality,'' {\em Physical Review E}, vol.~95, no.~3, p.~032407, 2017.

\bibitem{wu2009origin}
M.~Wu and P.~G. Higgs, ``Origin of self-replicating biopolymers: autocatalytic feedback can jump-start the rna world,'' {\em Journal of molecular evolution}, vol.~69, no.~5, pp.~541--554, 2009.

\bibitem{wu2012origin}
M.~Wu and P.~G. Higgs, ``The origin of life is a spatially localized stochastic transition,'' {\em Biology direct}, vol.~7, no.~1, pp.~1--15, 2012.

\bibitem{gillespie1976general}
D.~T. Gillespie, ``A general method for numerically simulating the stochastic time evolution of coupled chemical reactions,'' {\em Journal of computational physics}, vol.~22, no.~4, pp.~403--434, 1976.

\bibitem{erban2007practical}
R.~Erban, J.~Chapman, and P.~Maini, ``A practical guide to stochastic simulations of reaction-diffusion processes,'' {\em arXiv preprint arXiv:0704.1908}, 2007.

\bibitem{anderson2008incorporating}
D.~F. Anderson, ``Incorporating postleap checks in tau-leaping,'' {\em The Journal of chemical physics}, vol.~128, no.~5, p.~054103, 2008.

\bibitem{cao2006efficient}
Y.~Cao, D.~T. Gillespie, and L.~R. Petzold, ``Efficient step size selection for the tau-leaping simulation method,'' {\em The Journal of chemical physics}, vol.~124, no.~4, p.~044109, 2006.

\bibitem{frank1953spontaneous}
F.~C. Frank, ``On spontaneous asymmetric synthesis,'' {\em Biochimica et biophysica acta}, vol.~11, pp.~459--463, 1953.

\bibitem{daly2018ecological}
A.~J. Daly, J.~M. Baetens, and B.~De~Baets, ``Ecological diversity: measuring the unmeasurable,'' {\em Mathematics}, vol.~6, no.~7, p.~119, 2018.

\bibitem{onsager1944crystal}
L.~Onsager, ``Crystal statistics. i. a two-dimensional model with an order-disorder transition,'' {\em Physical Review}, vol.~65, no.~3-4, p.~117, 1944.

\bibitem{karlin1972application}
S.~Karlin and J.~McGregor, ``Application of method of small parameters to multi-niche population genetic models,'' {\em Theoretical Population Biology}, vol.~3, no.~2, pp.~186--209, 1972.

\bibitem{connell1978diversity}
J.~H. Connell, ``Diversity in tropical rain forests and coral reefs: high diversity of trees and corals is maintained only in a nonequilibrium state.,'' {\em Science}, vol.~199, no.~4335, pp.~1302--1310, 1978.

\bibitem{levin1998ecosystems}
S.~A. Levin, ``Ecosystems and the biosphere as complex adaptive systems,'' {\em Ecosystems}, vol.~1, no.~5, pp.~431--436, 1998.

\bibitem{gleason1927further}
H.~A. Gleason, ``Further views on the succession-concept,'' {\em Ecology}, vol.~8, no.~3, pp.~299--326, 1927.

\bibitem{levin2005self}
S.~A. Levin, ``Self-organization and the emergence of complexity in ecological systems,'' {\em Bioscience}, vol.~55, no.~12, pp.~1075--1079, 2005.

\bibitem{levin1985pattern}
S.~A. Levin and L.~A. Segel, ``Pattern generation in space and aspect,'' {\em SIAM Review}, vol.~27, no.~1, pp.~45--67, 1985.

\bibitem{cazzolla2020emergence}
R.~Cazzolla~Gatti, R.~Koppl, B.~D. Fath, S.~Kauffman, W.~Hordijk, and R.~E. Ulanowicz, ``On the emergence of ecological and economic niches,'' {\em Journal of Bioeconomics}, vol.~22, no.~2, pp.~99--127, 2020.

\bibitem{gatti2018niche}
R.~C. Gatti, B.~Fath, W.~Hordijk, S.~Kauffman, and R.~Ulanowicz, ``Niche emergence as an autocatalytic process in the evolution of ecosystems,'' {\em Journal of theoretical biology}, vol.~454, pp.~110--117, 2018.

\bibitem{godfrey2009darwinian}
P.~Godfrey-Smith, {\em Darwinian populations and natural selection}.
\newblock Oxford University Press, 2009.

\bibitem{hunding2006compositional}
A.~Hunding, F.~Kepes, D.~Lancet, A.~Minsky, V.~Norris, D.~Raine, K.~Sriram, and R.~Root-Bernstein, ``Compositional complementarity and prebiotic ecology in the origin of life,'' {\em Bioessays}, vol.~28, no.~4, pp.~399--412, 2006.

\bibitem{baum2015selection}
D.~A. Baum, ``Selection and the origin of cells,'' {\em Bioscience}, vol.~65, no.~7, pp.~678--684, 2015.

\bibitem{schlogl1972chemical}
F.~Schl{\"o}gl, ``Chemical reaction models for non-equilibrium phase transitions,'' {\em Zeitschrift f{\"u}r physik}, vol.~253, no.~2, pp.~147--161, 1972.

\bibitem{murray2001mathematical}
J.~D. Murray, {\em Mathematical biology II: spatial models and biomedical applications}, vol.~3.
\newblock Springer New York, 2001.

\bibitem{kerr2002local}
B.~Kerr, M.~A. Riley, M.~W. Feldman, and B.~J. Bohannan, ``Local dispersal promotes biodiversity in a real-life game of rock--paper--scissors,'' {\em Nature}, vol.~418, no.~6894, pp.~171--174, 2002.

\bibitem{munuzuri2022unified}
A.~P. Mu{\~n}uzuri and J.~P{\'e}rez-Mercader, ``Unified representation of life's basic properties by a 3-species stochastic cubic autocatalytic reaction-diffusion system of equations,'' {\em Physics of Life Reviews}, 2022.

\bibitem{gleiser2008extended}
M.~Gleiser and S.~I. Walker, ``An extended model for the evolution of prebiotic homochirality: A bottom-up approach to the origin of life,'' {\em Origins of Life and Evolution of Biospheres}, vol.~38, pp.~293--315, 2008.

\bibitem{gleiser2012life}
M.~Gleiser and S.~I. Walker, ``Life's chirality from prebiotic environments,'' {\em International Journal of Astrobiology}, vol.~11, no.~4, pp.~287--296, 2012.

\bibitem{shah2019survival}
V.~Shah, J.~de~Bouter, Q.~Pauli, A.~S. Tupper, and P.~G. Higgs, ``Survival of rna replicators is much easier in protocells than in surface-based, spatial systems,'' {\em Life}, vol.~9, no.~3, p.~65, 2019.

\bibitem{takeuchi2009multilevel}
N.~Takeuchi and P.~Hogeweg, ``Multilevel selection in models of prebiotic evolution ii: a direct comparison of compartmentalization and spatial self-organization,'' {\em PLoS computational biology}, vol.~5, no.~10, p.~e1000542, 2009.

\bibitem{luisi1989self}
P.~L. Luisi and F.~J. Varela, ``Self-replicating micelles—a chemical version of a minimal autopoietic system,'' {\em Origins of Life and Evolution of the Biosphere}, vol.~19, no.~6, pp.~633--643, 1989.

\bibitem{https://doi.org/10.21231/gnt1-hw21}
{Center for High Throughput Computing}, ``Center for high throughput computing,'' 2006.

\end{thebibliography}

\end{document}